\documentclass[10pt,twoside]{cras_iv-en}
\usepackage{amssymb,amsbsy,amsmath,amsfonts,amssymb,amscd}
\usepackage{latexsym,euscript,exscale,epsfig}
\usepackage[english,francais]{babel}
\newtheorem{e-proposition}[theorem]{Proposition}

\newtheorem{e-definition}[theorem]{D\'efinition\rm}

\ComParit{S1296-2147}
\PIT{FLA}
\PXHY{1171}
\Add{?}
\Volume{2}
\Year{2001}
\FirstPage{1}
\LastPage{??}
\AuteurCourant{Luc Blanchet}
\TitreCourant{On the two-body problem in general relativity} 
\Journal
\Rubrique{Rubrique}{Heading}
\SousRubrique{Sous-rubrique}{Sub-Heading}
\PresentePar{First name}{NAME}
\Recu{jour mois ann\'ee}{apr\`es r\'evision}{jour mois ann\'ee}
\motscles{relativit\'e g\'en\'erale/\'equations du mouvement}
\begin{document}
\selectlanguage{english}
\TitleOfDossier{
}
\TitreDeDossier{
}
\title{On the two-body problem in general relativity
}
\author{Luc Blanchet
$^{\text{a}}$
}
\address{
\begin{itemize}\labelsep=2mm\leftskip=-5mm
\item[$^{\text{a}}$]
Institut d'Astrophysique de Paris, 98bis boulevard
Arago, 75014 Paris, France\\
E-mail: blanchet@iap.fr
\end{itemize}
}
\maketitle
\thispagestyle{empty}
\begin{Abstract}{
We consider the two-body problem in post-Newtonian approximations of
general relativity. We report the recent results concerning the
equations of motion, and the associated Lagrangian formulation, of
compact binary systems, at the third post-Newtonian order ($\sim
1/c^6$ beyond the Newtonian acceleration). These equations are
necessary when constructing the theoretical templates for searching and
analyzing the gravitational-wave signals from inspiralling compact
binaries in VIRGO-type experiments.  }\end{Abstract}
\selectlanguage{french}
\begin{Ftitle}{
Sur le probl\`eme des deux corps en relativit\'e g\'en\'erale
}\end{Ftitle}
\begin{Resume}{
~~Nous consid\'erons le probl\`eme des deux corps dans l'approximation
post-newtonien-ne de la relativit\'e g\'en\'erale. Nous pr\'esentons
les r\'esultats r\'ecents concernant les \'equations du mouvement, et
la formulation lagrangienne associ\'ee, de syst\`emes binaires
d'objets compacts, au troisi\`eme ordre post-newtonien ($\sim 1/c^6$
apr\`es l'acc\'e-l\'eration newtonienne). Ces \'equations sont
utilis\'ees dans la construction des filtres th\'eoriques pour la
d\'etection et l'analyse des signaux d'ondes gravitationnelles
provenant des binaires compactes spiralantes dans les exp\'eriences du
type VIRGO.  }\end{Resume}


\par\medskip\centerline{\rule{2cm}{0.2mm}}\medskip
\setcounter{section}{0}
\selectlanguage{english}
By two-body problem we mean the problem of the dynamics of two
strutureless, non-spinning point-particles, characterized by solely
two mass parameters $m_1$ and $m_2$, moving under their mutual, purely
gravitational, interaction. Surely this problem, because of its
conceptual simplicity, is among the most interesting ones to be solved
within any theory of gravity. Actually, there are two aspects of the
problem: the first sub-problem consists of obtaining the {\it
equation} of the binary motion, the second is to find the (hopefully
exact) solution of that equation. We refer to the equation of motion
as the explicit expression of the acceleration of each of the
particles in terms of their positions and velocities. It is well known
that in Newtonian gravity, the first of these sub-problems is trivial,
as one can easily write down the equation of motion for a system of
$N$ particles, while the second one is difficult, except in the
two-body case $N=2$, which represents in fact the only situation
amenable to an exact treatment of the solution.

In general relativity, even writing down the equations of motion in the
simplest case $N=2$ is difficult. Unlike in Newton's theory, it is
impossible to express the acceleration by means of the positions and
velocities, in a way which would be valid within the ``exact''
theory. Therefore we are obliged to resort to approximation
methods. Let us feel reassured that plaguing the exact theory of
general relativity with approximation methods is not a shame. It is
fair to say that many of the great successes of this theory, when
confronted to experiments and observations, have been obtained thanks
to approximation methods. Furthermore, the beautiful internal wheels
of general relativity {\it also} show up when using approximation
methods, which often deserve some theoretical interest in their own,
as they require interesting mathematical techniques.

Here we whall investigate the equation of the binary motion in the
post-Newtonian approximation, i.e. as a formal expansion when the
velocity of light $c$ tends to infinity, up to the so-called third
post-Newtonian (3PN) order, i.e. $1/c^6$ beyond the usual Newtonian
acceleration. This problem is not pure academicism, because the
equation of motion at this somewhat frightening 3PN order is needed
(among other things) during the computation of the orbital phase of
the inspiralling compact binaries \cite{BIJ01,BFIJ01}. The phase
constitutes the crucial component of the theoretical templates to be
used for searching and analyzing the gravitational waves from the
binary inspiral in VIRGO-type detectors.

We write the 3PN acceleration of the particle 1, say, in the form

\begin{equation}\label{1}
a_1^i = A_{\rm N}^i + \frac{1}{c^2}A_{\rm 1PN}^i +
\frac{1}{c^4}A_{\rm 2PN}^i + \frac{1}{c^5}A_{\rm 2.5PN}^i +
\frac{1}{c^6}A_{\rm 3PN}^i + {\cal O}\left(\frac{1}{c^7}\right)\;,
\end{equation}
where the first term is given by the famous Newtonian law,

\begin{equation}\label{2}
A_{\rm N}^i = -\frac{G m_2}{r_{12}^2}n_{12}^i \;.
\end{equation}
For simplicity, we do not write the particle's label 1 on the
$A_{\rm nPN}^i$'s. We denote by $r_{12}=|{\bf y}_1(t)-{\bf y}_2(t)|$ the
distance between the two particles, with ${\bf y}_1=(y_1^i)$ and ${\bf
y}_2=(y_2^i)$ their ``absolute'' positions, by
$n_{12}^i=(y_1^i-y_2^i)/r_{12}$ the corresponding unit direction, and
by $a_1^i=dv_1^i/dt$ and $v_1^i=dy_1^i/dt$ the acceleration and
velocity of 1 (and {\it idem} for 2). Sometimes we pose
$v_{12}^i=v_1^i-v_2^i$ for the relative velocity; and similarly for
the relative position and acceleration. The equation for the other
body is obtained by exchanging the labels $1\leftrightarrow 2$
(remembering that $n_{12}^i$ changes sign in this operation).

As a consequence of the equivalence principle, which is incorporated
``by hand'' in Newton's theory and constitutes the fundamental basis
of general relativity, the acceleration of 1 should not depend on
$m_1$ (nor on its internal structure), in the ``test-mass'' limit
where the mass $m_1$ is much smaller than $m_2$. This is of course
satisfied by the Newtonian acceleration (\ref{2}), which is
independent of $m_1$, but this leaves the possibility that the
acceleration of the particle 1, in higher approximations, does depend
on $m_1$, via the so-called self-forces, which vanish in the test-mass
limit. Indeed, this is what happens in the post-Newtonian corrections
computed in Eqs. (\ref{3})-(\ref{6}), which show explicitly many terms
proportional to (powers of) $m_1$.
 
Though the successive post-Newtonian approximations are really a
consequence of general relativity, they should be interpreted using
the common-sense language of Newton. That is, having chosen a
convenient general-relativistic (Cartesian) coordinate system, like
the harmonic coordinate system adopted below, we should express the
results in terms of the {\it coordinate} positions, velocities and
accelerations of the bodies. Then, the trajectories of the particles
can be viewed as taking place in the absolute Euclidean space of
Newton, and their (coordinate) velocities as being defined with
respect to absolute time. Not only this interpretation is the most
satisfactory one from a conceptual point of view, but it represents
also the most convenient path for comparing the theoretical
predictions and the observations. For instance, the solar system
dynamics at the first post-Newtonain level is defined, following a
recent resolution of the International Astronomical Union, in a
harmonic coordinate system, the Geocentric Reference System (GRS),
with respect to which one considers the ``absolute'' motion of the
planets and satellites.

With the same spirit we present below the 3PN equations of motion in
terms of the harmonic-coordinate positions and velocities of the
particles, i.e., in particular, in a form which is not {\it
manifestly} covariant. But because the equations come from general
relativity, they are endowed with the following properties, which make
them truly ``relativistic''.

\medskip\noindent (I) The one-body problem in general relativity
corresponds to the Schwarzschild solution, so the equations possess
the correct ``perturbative'' limit, that given by the geodesics of the
Schwarzschild metric (developed to the 3PN order), when the mass of
one of the bodies tends to zero.

\medskip\noindent (II) Because general relativity admits the
Poincar\'e group as a global symmetry (in the case of asymptotically
flat space-times), the harmonic-coordinate equations of motion stay
invariant when we perform a global Lorentz transformation, expanded at
the 3PN order.

\medskip\noindent (III) Since the particles emit gravitational
radiation there are some terms in the equations which are associated
with radiation reaction. These terms appear at the order 2.5PN or
$1/c^5$ (where $5=2s+1$, $s=2$ being the helicity of the graviton);
see Eq. (\ref{1}). They correspond to an ``odd''-order post-Newtonian
correction, which does not stay invariant in a time reversal. By
contrast, as we shall see, the ``even''-orders 1PN, 2PN and 3PN
correspond to a dynamics which is conservative.

\medskip\noindent (IV) General relativity is a non linear theory (even
in vacuum), and some part of the gravitational radiation which was
emitted by the particles in the past scatters off the static
gravitational field generated by the rest-masses of the particles, or
interacts gravitationally with itself. The ``tail'' radiation,
produced by non-linear scattering, converges back onto the system at
our present epoch, and modifies its current dynamics. The effect
appears at the 4PN order [negligible in Eq. (\ref{1})], and
corresponds to some ``even''-order modification of the
radiation-reaction damping \cite{BD88}.

\medskip
The dominant relativistic correction term (1PN, or $1/c^2$) in the
case of two non-spinning compact bodies was derived first by Lorentz
and Droste \cite{LD17}. Subsequently, Einstein, Infeld and Hoffmann
\cite{EIH} obtained the 1PN corrections by means of their famous
``surface-integral'' method, in which the equations of motion are
deduced from the {\it vacuum} field equations. This method is
applicable to any compact objects (they be neutron stars, black holes,
or, maybe, naked singularities). The 1PN-accurate equations were also
obtained, for the motion of the centers of mass of extended bodies, by
Fock and followers \cite{Fock}, using a technique that can be
qualified as more ``physical'' than the surface-integral method, as it
takes explicitly into account the internal structure of the
bodies. The result is

\begin{eqnarray}\label{3}
A_{\rm 1PN}^i &=& \bigg[5\frac{G^2 m_1 m_2}{r_{12}^3} + 4\frac{
G^2 m_2^2}{r_{12}^3 }+ \frac{G m_2}{r_{12}^2} \bigg(\frac{3}{2}
(n_{12}v_2)^2 - v_1^2 + 4 (v_1v_2) - 2 v_2^2\bigg)\bigg] n_{12}^i
\nonumber\\ &+& \frac{G m_2}{r_{12}^2} \bigg[4 (n_{12}v_1) -
3 (n_{12}v_2)\bigg] v_{12}^i\;. 
\end{eqnarray}
We denote the usual Euclidean scalar product by parenthesis, for
instance $(n_{12}v_1)={\bf n}_{12}.{\bf v}_1$ and $(v_1v_2)={\bf
v}_1.{\bf v}_2$. Witness the first term in Eq. (\ref{3}), proportional
to $m_1$, which represents a self-force at the 1PN order. The
1PN-accurate equations of motion (known also for $N$ compact bodies)
are sometimes called after the names of Einstein, Infeld and Hoffmann
\cite{EIH}.

Things become more and more difficult when going to higher and higher
post-Newtonian approximations. A technical problem is the rapid
proliferation of terms. Typically any allowed term (compatible
dimension, correct mass dependence) does appear with a definite
non-zero coefficient in front. As a result, the expressions look quite
complicated; that's maybe the price we have to pay for expressing in a
Newtonian fashion what really is a relativistic equation. But, the
point for our purpose is that the Newtonian-like equations are fully
explicit: the successive relativistic corrections depend only on the
instantaneous positions and velocities --- all accelerations,
appearing at intermediate stages of the calculation, are consistently
``order-reduced'' by means of the equations of motion themselves.
 
The 2PN approximation was tackled by Otha {\it et al} \cite{OO}, how
considered the post-Newtonian iteration of the Hamiltonian of $N$
point-particles. We refer here to the Hamiltonian as the Fokker-type
Hamiltonian, which is obtained from the matter-plus-field
Arnowitt-Deser-Misner (ADM) Hamiltonian by eliminating the field
degrees of freedom. The result for the 2PN equations of binary
motion in harmonic coordinates was obtained by Damour and Deruelle
\cite{DD,D1,D2}, building on a non-linear iteration of the metric of two
particles initiated in Ref. \cite{BeDD81}. It reads

\begin{eqnarray} \label{4}
A_{\rm 2PN}^i &=& \bigg[-\frac{57}{4}\frac{G^3 m_1^2 m_2}{r_{12}^4}-
\frac{69}{2}\frac{G^3 m_1 m_2^2}{r_{12}^4}- 9\frac{G^3 m_2^3}{r_{12}^4} +
\frac{G m_2}{r_{12}^2} \bigg(-\frac{15}{8} (n_{12}v_2)^4 + \frac{3}{2}
(n_{12}v_2)^2 v_1^2 \nonumber\\ &-& 6 (n_{12}v_2)^2 (v_1v_2)
- 2 (v_1v_2)^2 + \frac{9}{2} (n_{12}v_2)^2 v_2^2 + 4 (v_1v_2) v_2^2 -
2 v_2^4\bigg) \nonumber\\ &+& \frac{G^2 m_1 m_2}{ r_{12}^3}
\bigg(\frac{39}{2} (n_{12}v_1)^2 - 39 (n_{12}v_1) (n_{12}v_2) +
\frac{17}{2} (n_{12}v_2)^2 - \frac{15}{4} v_1^2 - \frac{5}{2} (v_1v_2)
+ \frac{5}{4} v_2^2\bigg) \nonumber\\ &+& \frac{G^2
m_2^2}{r_{12}^3} \bigg(2 (n_{12}v_1)^2 - 4 (n_{12}v_1) (n_{12}v_2) - 6
(n_{12}v_2)^2 - 8 (v_1v_2) + 4 v_2^2\bigg)\bigg] n_{12}^i \nonumber\\
&+& \bigg[\frac{G^2 m_2^2}{r_{12}^3} \bigg(-2 (n_{12}v_1) - 2
(n_{12}v_2)\bigg) + \frac{G^2 m_1 m_2}{r_{12}^3} \bigg(-\frac{63}{4}
(n_{12}v_1) + \frac{55}{4} (n_{12}v_2)\bigg) \nonumber\\ &+&
\frac{G m_2}{r_{12}^2} \bigg(-6 (n_{12}v_1) (n_{12}v_2)^2 +
\frac{9}{2} (n_{12}v_2)^3 + (n_{12}v_2) v_1^2 - 4 (n_{12}v_1) (v_1v_2)
\nonumber\\ &+& 4 (n_{12}v_2) (v_1v_2) + 4 (n_{12}v_1) v_2^2
- 5 (n_{12}v_2) v_2^2\bigg)\bigg] v_{12}^i\;.
\end{eqnarray}
The corresponding result for the ADM-Hamiltonian of two particles at
the 2PN order was given by Damour and Sch\"afer \cite{DS85}.  

As mentioned above, the equation of motion is conservative up to the
2PN level. Only at the 2.5PN order appears the first non-conservative
effect, associated with the gravitational radiation reaction. The
(harmonic-coordinate) equation of motion up to that level has been
derived by Damour and Deruelle \cite{DD,D1,D2}, who used it for the
study of the radiation damping of the orbital period of the binary
pulsar PSR~1913+16 \cite{D2}. In addition, the 2.5PN-accurate
ADM-Hamiltonian was worked out by Sch\"afer \cite{S}, and the
harmonic-coordinate equations as well as the complete gravitational
field of two particles were computed by Blanchet, Faye and Ponsot
\cite{BFP98} by means of a new method coming from the work on
gravitational-wave generation \cite{BIJ01}. The latter calculations
use a formal description of the objects by delta functions.  Needless
to say, there is mutual agreement between all the results obtained so
far. The 2.5PN radiation damping term reads

\begin{eqnarray} \label{5}
A_{\rm 2.5PN}^i &=& \frac{4}{5}\frac{G^2 m_1 m_2}{r_{12}^3} \bigg[ -
6\frac{G m_1}{r_{12}} + \frac{52}{3}\frac{G m_2}{r_{12}} + 3v_{12}^2
\bigg] (n_{12}v_{12}) n_{12}^i \nonumber\\ &+& \frac{4}{5}\frac{G^2 m_1
m_2}{r_{12}^3}\bigg[2\frac{G m_1}{r_{12}} - 8\frac{G m_2}{r_{12}} -
v_{12}^2 \bigg] v_{12}^i\;.
\end{eqnarray}
It is important to realize that the 2.5PN equations of motion
(\ref{1})-(\ref{5}) have been proved to hold in the case of binary
systems of strongly self-gravitating bodies \cite{D2}. This is {\it
via} an effacing principle of the internal structure of the bodies. As
a result, the equations depend only on the ``Schwarzschild'' masses,
$m_1$ and $m_2$, of the compact objects. Compacity parameters, namely
the ratios $\frac{G m_1}{b_1c^2}$ and $\frac{G m_2}{b_2c^2}$ between
the masses and the radii $b_1$ and $b_2$ of the objects, do not enter
the equations of motion. This fact has been explicitly verified at the
2.5PN order by Grishchuk and Kopejkin \cite{Kop}, who have made a
``physical'' computation, {\it \`a la} Fock, taking into account the
internal structure of two self-gravitating extended
bodies. Furthermore, the same 2.5PN equations of motion have also been
established by Itoh, Futamase and Asada \cite{IFA}, who use a variant
of the surface-integral approach \cite{EIH}, valid for
compact bodies, independently of the strength of the internal
gravity.

The present state of the art is the 3PN approximation. The equations
to this order have been worked out independently by two groups, by
means of different methods, and with equivalent results. On one hand,
Jaranowski and Sch\"afer \cite{JaraS}, and Damour, Jaranowski and
Sch\"afer \cite{DJS}, following the line of research of
Refs. \cite{OO,DS85,S}, employ the ADM-Hamiltonian formalism of
general relativity; on the other hand, Blanchet and Faye
\cite{BF,BFreg}, and Andrade, Blanchet and Faye \cite{ABF}, founding
their approach on the post-Newtonian iteration initiated in
Ref. \cite{BFP98}, compute directly the harmonic-coordinate equations
of motion. The end results are physically equivalent in the sense that
there exists a unique ``contact'' transformation of the dynamical
variables, that changes the harmonic-coordinate Lagrangian obtained in
Ref. \cite{ABF} into a new Lagrangian, whose associated Hamiltonian
coincides exactly with the one given in Ref. \cite{DJS}. The 3PN
contribution to the acceleration of the particle 1 reads, in a
Cartesian harmonic coordinate system,

\begin{eqnarray} \label{6}
A_{\rm 3PN}^i &=& \bigg[\frac{G m_2}{r_{12}^2} \bigg(\frac{35}{16}
(n_{12}v_2)^6 - \frac{15}{8} (n_{12}v_2)^4 v_1^2 + \frac{15}{2}
(n_{12}v_2)^4 (v_1v_2) + 3 (n_{12}v_2)^2 (v_1v_2)^2 \nonumber\\ &-&
\frac{15}{2} (n_{12}v_2)^4 v_2^2 + \frac{3}{2} (n_{12}v_2)^2 v_1^2
v_2^2 - 12 (n_{12}v_2)^2 (v_1v_2) v_2^2 - 2 (v_1v_2)^2 v_2^2
\nonumber\\ &+& \frac{15}{2} (n_{12}v_2)^2 v_2^4 + 4 (v_1v_2) v_2^4 -
2 v_2^6\bigg) + \frac{G^2 m_1 m_2}{ r_{12}^3 } \bigg(-\frac{171}{8}
(n_{12}v_1)^4 \nonumber\\ &+& \frac{171}{2} (n_{12}v_1)^3 (n_{12}v_2)
- \frac{723}{4} (n_{12}v_1)^2 (n_{12}v_2)^2 + \frac{383}{2}
(n_{12}v_1) (n_{12}v_2)^3 \nonumber\\
&-& \frac{455}{8} (n_{12}v_2)^4
+ \frac{229}{4} (n_{12}v_1)^2 v_1^2 - \frac{205}{2} (n_{12}v_1)
(n_{12}v_2) v_1^2 + \frac{191}{4} (n_{12}v_2)^2 v_1^2 - \frac{91}{8}
v_1^4 \nonumber\end{eqnarray}
\begin{eqnarray} 
\qquad &-& \frac{229}{2} (n_{12}v_1)^2 (v_1v_2) + 244 (n_{12}v_1)
(n_{12}v_2) (v_1v_2) - \frac{225}{2} (n_{12}v_2)^2 (v_1v_2)\nonumber\\
&+& \frac{91}{2} v_1^2 (v_1v_2) - \frac{177}{4} (v_1v_2)^2 +
\frac{229}{4} (n_{12}v_1)^2 v_2^2 - \frac{283}{2} (n_{12}v_1)
(n_{12}v_2) v_2^2 \nonumber\\ &+& \frac{259}{4} (n_{12}v_2)^2 v_2^2 -
\frac{91}{4} v_1^2 v_2^2 + 43 (v_1v_2) v_2^2 - \frac{81}{8}
v_2^4\bigg) + \frac{G^2 m_2^2}{ r_{12}^3 } \bigg(-6 (n_{12}v_1)^2
(n_{12}v_2)^2 \nonumber\\ &+& 12 (n_{12}v_1) (n_{12}v_2)^3 + 6
(n_{12}v_2)^4 + 4 (n_{12}v_1) (n_{12}v_2) (v_1v_2) + 12 (n_{12}v_2)^2
(v_1v_2)\nonumber\\ &+& 4 (v_1v_2)^2 - 4 (n_{12}v_1) (n_{12}v_2) v_2^2
- 12 (n_{12}v_2)^2 v_2^2 - 8 (v_1v_2) v_2^2 + 4 v_2^4\bigg)
\nonumber\\ &+& \frac{G^3 m_2^3}{r_{12}^4} \bigg(-(n_{12}v_1)^2 + 2
(n_{12}v_1) (n_{12}v_2) + \frac{43}{2} (n_{12}v_2)^2 + 18 (v_1v_2) - 9
v_2^2\bigg) \nonumber\\ &+& \frac{G^3 m_1 m_2^2}{r_{12}^4}
\bigg(\frac{415}{8} (n_{12}v_1)^2 - \frac{375}{4} (n_{12}v_1)
(n_{12}v_2) + \frac{1113}{8} (n_{12}v_2)^2 \nonumber\\ &-&
\frac{615}{64}(n_{12}v_{12})^2 \pi^2 + \frac{123}{64} v_{12}^2\pi^2 +
18 v_1^2 + 33 (v_1v_2) - \frac{33}{2} v_2^2 \bigg) \nonumber\\ &+&
\frac{G^3 m_1^2 m_2}{r_{12}^4} \bigg(-\frac{45887}{168} (n_{12}v_1)^2
+ \frac{24025}{42} (n_{12}v_1) (n_{12}v_2) - \frac{10469}{42}
(n_{12}v_2)^2 \nonumber\\ &+& \frac{48197}{840} v_1^2 -
\frac{36227}{420} (v_1v_2) + \frac{36227}{840} v_2^2 + 110
(n_{12}v_{12})^2 \ln \left(\frac{r_{12}}{r'_1} \right) - 22 v_{12}^2
\ln \left(\frac{r_{12}}{r'_1} \right) \bigg) \nonumber\\ &+& \frac{G^4
m_1^3 m_2}{r_{12}^5} \bigg(-\frac{3187}{1260} + \frac{44}{3} \ln
\left(\frac{r_{12}}{r'_1} \right)\bigg) + \frac{G^4 m_1^2
m_2^2}{r_{12}^5} \bigg(\frac{34763}{210} - \frac{44}{3}\lambda -
\frac{41}{16} \pi^2\bigg) \nonumber\\ &+& \frac{G^4 m_1
m_2^3}{r_{12}^5} \bigg(\frac{10478}{63} - \frac{44}{3}\lambda -
\frac{41}{16} \pi^2 - \frac{44}{3} \ln \left(\frac{r_{12}}{r'_2}
\right)\bigg)+16\frac{G^4 m_2^4}{r_{12}^5} \bigg] n_{12}^i \nonumber\\
&+& \bigg[\frac{G m_2}{r_{12}^2} \bigg(\frac{15}{2} (n_{12}v_1)
(n_{12}v_2)^4 - \frac{45}{8} (n_{12}v_2)^5 - \frac{3}{2} (n_{12}v_2)^3
v_1^2 + 6 (n_{12}v_1) (n_{12}v_2)^2 (v_1v_2) \nonumber\\ &-& 6
(n_{12}v_2)^3 (v_1v_2) - 2 (n_{12}v_2) (v_1v_2)^2 - 12
(n_{12}v_1)(n_{12}v_2)^2 v_2^2 + 12 (n_{12}v_2)^3 v_2^2 \nonumber\\
&+& (n_{12}v_2) v_1^2 v_2^2 - 4 (n_{12}v_1) (v_1v_2) v_2^2 + 8
(n_{12}v_2) (v_1v_2) v_2^2 + 4 (n_{12}v_1) v_2^4 - 7 (n_{12}v_2)
v_2^4\bigg) \nonumber\\ &+& \frac{G^2 m_2^2}{r_{12}^3} \bigg(-2
(n_{12}v_1)^2 (n_{12}v_2) + 8 (n_{12}v_1) (n_{12}v_2)^2 + 2
(n_{12}v_2)^3 \nonumber\\ &+& 2 (n_{12}v_1) (v_1v_2) + 4 (n_{12}v_2)
(v_1v_2) - 2 (n_{12}v_1) v_2^2 - 4 (n_{12}v_2) v_2^2\bigg) \nonumber\\
&+& \frac{G^2 m_1 m_2}{r_{12}^3} \bigg(-\frac{243}{4} (n_{12}v_1)^3 +
\frac{565}{4} (n_{12}v_1)^2 (n_{12}v_2) - \frac{269}{4} (n_{12}v_1)
(n_{12}v_2)^2 \nonumber\\ &-& \frac{95}{12} (n_{12}v_2)^3 +
\frac{207}{8} (n_{12}v_1) v_1^2 - \frac{137}{8} (n_{12}v_2) v_1^2 - 36
(n_{12}v_1) (v_1v_2) \nonumber\\ &+& \frac{27}{4} (n_{12}v_2)
(v_1v_2)+ \frac{81}{8} (n_{12}v_1) v_2^2 + \frac{83}{8} (n_{12}v_2)
v_2^2\bigg) + \frac{G^3 m_2^3}{r_{12}^4} \bigg(4 (n_{12}v_1) + 5
(n_{12}v_2)\bigg) \nonumber\\ &+& \frac{G^3 m_1 m_2^2}{r_{12}^4}
\bigg(-\frac{307}{8} (n_{12}v_1) + \frac{479}{8} (n_{12}v_2) +
\frac{123}{32} (n_{12}v_{12}) \pi^2 \bigg) \nonumber\\ &+& \frac{G^3
m_1^2 m_2}{r_{12}^4} \bigg(\frac{31397}{420} (n_{12}v_1) -
\frac{36227}{420} (n_{12}v_2) - 44 (n_{12}v_{12}) \ln
\left(\frac{r_{12}}{r'_1} \right) \bigg)\bigg] v_{12}^i \;.
\end{eqnarray}
Notice that the 3PN term involves some $\pi^2$. Typically, the $\pi^2$
terms arise from non-linear interactions involving some integrals such
as $\frac{1}{\pi}\int \frac{d^3{\bf
x}}{r_1^2r_2^2}=\frac{\pi^2}{r_{12}}$.

More importantly, we see that the 3PN acceleration depends on three
arbitrary constants: two length scales $r'_1$ and $r'_2$ entering some
logarithms, and a dimensionless constant $\lambda$. It was proved in
Ref. \cite{BF} that $r'_1$ and $r'_2$ are merely linked with the
choice of harmonic coordinates. Indeed, as we are using
point-particles, the usual condition of harmonic coordinates, {\it
viz} $\partial_\nu h^{\mu\nu}=0$, does not completely fix the gauge,
as we can always add a gauge vector $\xi^\mu=\delta x^\mu$, which
satisfies $\Delta\xi^\mu=0$ and is singular at the location of the two
particles. The constants $r'_1$ and $r'_2$ can thus be removed by a
coordinate transformation; as a result they will never appear in any
physical result, such as the invariant center-of-mass energy of the
binary.
 
By contrast with the harmless gauge-constants $r'_1$ and $r'_2$, the
constant $\lambda$ represents a true physical ambiguity, which
reflects probably an incompleteness of the Hadamard regularization
used to cope with the infinite self-field of point-particles. This
regularization is based on Hadamard's concept of the ``partie finie''
of singular functions and divergent integrals. Actually, it has been
found necessary for solving this problem to develop an extended
version of the Hadamard regularization, and a theory of generalized
functions associated with it \cite{BFreg}. Very likely, the presence
of $\lambda$ is related to the fact that, starting from the 3PN order,
many non-linear integrals composing the equations of motion, when taken
separately, depend on the internal structure of each body, even in the
limit where the radius tends to zero. The results given by the
ADM-Hamiltonian approach \cite{JaraS,DJS} depend also on one arbitrary
physical parameter, called $\omega_{\rm static}$. More precisely, the
authors \cite{JaraS} introduced originally two unknown constants,
$\omega_{\rm static}$ and $\omega_{\rm kinetic}$, but $\omega_{\rm
kinetic}$ was fixed later to a unique value by imposing, in an {\it ad
hoc} manner, the global Poincar\'e invariance of the Hamiltonian
\cite{DJS}. On the other hand, the authors \cite{BF,BFreg} have only
one constant $\lambda$ because their regularization is defined in a
Lorentz-invariant way. The equivalence between the approaches
\cite{BF,BFreg,ABF} and \cite{JaraS,DJS} holds if and only if
$\lambda=-\frac{3}{11}\omega_{\rm
static}-\frac{1987}{3080}$. Recently, the value $\omega_{\rm
static}=0$ has been obtained by means of a different regularization
(dimensional) within the ADM-Hamiltonian approach \cite{DJS01}. This
result would mean that $\lambda=-\frac{1987}{3080}$ (but we keep
$\lambda$ unspecified in the present discussion).

Going to still higher post-Newtonian orders, we mention that the
subdominant radiation-reaction effect at the 3.5PN order has been
calculated in Ref. \cite{IW} for two bodies in an arbitrary gauge, and
in Ref. \cite{B93} for general ``fluids'' in a Burke-Thorne-extended
gauge (there is agreement between the two methods \cite{IW}). Also, we
know the contribution of the tails of waves in the equations of
motion, which appears at the 4PN, or $1/c^8$, order \cite{BD88}.

The complicated expressions (\ref{1})-(\ref{6}) simplify drastically
in the case of an orbit which is circular, apart from the gradual
inspiral driven by radiation reaction. This case corresponds to the
physical situation of most inspiralling compact binaries, since the
radiation reaction forces tend to circularize rapidly the orbit. In
this case the relative acceleration reads

\begin{equation}\label{7}
a_{12}^i = -\omega^2 y_{12}^i-
\frac{32}{5}\frac{G^3m^3\nu}{c^5r_{12}^4}v_{12}^i + {\cal
O}\left(\frac{1}{c^7}\right)\;.
\end{equation}
Relative quantities are denoted e.g. by $y_{12}^i=y_1^i-y_2^i$; mass
parameters are the total mass $m=m_1+m_2$ and the mass ratio
$\nu=m_1m_2/m^2$ such that $0 < \nu\leq \frac{1}{4}$, with
$\nu=\frac{1}{4}$ when the two masses are equal, and $\nu\to 0$ in the
test-mass limit for one of the particles. The second term in
Eq. (\ref{7}) is the radiation reaction force, whose expression
follows immediately from Eq. (\ref{5}), while $\omega$ represents the
orbital frequency of the circular motion at the 3PN order, and is
related to the harmonic-coordinate separation $r_{12}$ by the
``Kepler'' law

\begin{eqnarray}\label{8}
\omega^2 &=& \frac{Gm}{r_{12}^3} \bigg\{ 1+(-3+\nu) \gamma + \left( 6
+\frac{41}{4} \nu +\nu^2 \right) \gamma^2 \nonumber\\ &+& \left( -10 +
\left[-\frac{67759}{840}+\frac{41}{64}\pi^2
+22\ln\left(\frac{r}{r'_0}\right)+\frac{44}{3}\lambda\right]\nu
+\frac{19}{2}\nu^2+\nu^3 \right) \gamma^3 \bigg\}\;.
\end{eqnarray}
The post-Newtonian corrections are parametrized by
$\gamma=\frac{Gm}{r_{12}c^2}$. The constant $r'_0$ is given by $\ln
r'_0=\frac{m_1}{m}\ln r'_1+\frac{m_2}{m}\ln r'_2$, where $r'_1$ and
$r'_2$ are the two gauge-constants in Eq. (\ref{6}). The
circular-orbit equation (\ref{7})-(\ref{8}) is used, notably, to
compute the third time-derivative of the binary's quadrupole moment,
which constitutes the main contribution to the gravitational-wave flux
at infinity, at the 3PN order \cite{BIJ01}.

Finally, coming back to the general case of non-circular orbits, we
show that the 3PN equations of motion, when neglecting the
radiation-reaction term at the 2.5PN order, admit a Euler-Lagrange
formulation. As it turns out, the Lagrangian in harmonic coordinates
is a generalized one, in the sense that it depends not only on the
positions and velocities of the particles, like an ordinary
Lagrangian, but also, starting at the 2PN order, on their
accelerations. This fact has been established in
Refs. \cite{DD,D2,DS85}. Thus, the equations of motion take the form

\begin{equation}\label{9}
\frac{\partial L}{\partial y_1^i}-\frac{d}{dt}\left(\frac{\partial
L}{\partial v_1^i}\right) +\frac{d^2}{dt^2}\left(\frac{\partial
L}{\partial a_1^i}\right) + \frac{1}{c^5} m_1 A_{\rm 2.5PN}^i = 
{\cal O}\left(\frac{1}{c^7}\right)\;,
\end{equation}
where $A_{\rm 2.5PN}^i$ is given by Eq. (\ref{5}), and where the 3PN
harmonic-coordinate generalized Lagrangian is of the type

\begin{equation}\label{10}
L = L_{\rm N} + \frac{1}{c^2}L_{\rm 1PN} + \frac{1}{c^4}L_{\rm 2PN} +
\frac{1}{c^6}L_{\rm 3PN} + {\cal O}\left(\frac{1}{c^8}\right)\;.
\end{equation}
The Newtonian piece reads

\begin{eqnarray}\label{11}
L_{\rm N} &=& \frac{m_1 v_1^2}{2} + \frac{m_2 v_2^2}{2} + \frac{G m_1
m_2}{r_{12}}\;.
\end{eqnarray}
The 1PN piece, which is still ``ordinary'' (not depending on
accelerations), has been obtained by Lorentz and Droste \cite{LD17},
and Fichtenholz \cite{Fich}:

\begin{eqnarray}\label{12}
L_{\rm 1PN} &=& \frac{m_1
v_1^4}{8 }+ \frac{G m_1 m_2}{r_{12}} \bigg(-\frac{1}{4} (n_{12}v_1)
(n_{12}v_2) + \frac{3}{2} v_1^2 - \frac{7}{4} (v_1v_2)\bigg)\nonumber\\
&-& \frac{G^2 m_1^2 m_2}{2 r_{12}^2} + 1 \leftrightarrow 2\;.
\end{eqnarray}
To the expression given above, one must add the terms corresponding to
the label exchange $1\leftrightarrow 2$, including those that are
already symmetric under exchange. Now, starting at the 2PN order, we
get a dependence over the accelerations \cite{DD,D2,DS85}. By adding to
the Lagrangian some so-called ``multi-zero'' terms, which do not
contribute to the equations of motion, one can always arrange that the
dependence over the accelerations be {\it linear}. Of course, it is
not allowed to replace the accelerations by the equations of motion in
a Lagrangian (however, this can and should be done in the final
expressions of the conserved integrals derived from that
Lagrangian). We get

\begin{eqnarray}\label{13}
L_{\rm 2PN} &=& \frac{m_1 v_1^6}{16}+\frac{1}{2}\frac{G^3 m_1^3
m_2}{r_{12}^3}+ \frac{19}{8}\frac{G^3 m_1^2 m_2^2}{r_{12}^3} +
\frac{G^2 m_1^2 m_2}{r_{12}^2} \bigg(\frac{7}{2} (n_{12}v_1)^2 -
\frac{7}{2} (n_{12}v_1) (n_{12}v_2) \nonumber\\ &+&
\frac{1}{2}(n_{12}v_2)^2 + \frac{1}{4} v_1^2 - \frac{7}{4} (v_1v_2) +
\frac{7}{4} v_2^2\bigg) + \frac{G m_1 m_2}{r_{12}} \bigg(\frac{3}{16}
(n_{12}v_1)^2 (n_{12}v_2)^2 \nonumber\\ &-& \frac{7}{8} (n_{12}v_2)^2
v_1^2 + \frac{7}{8} v_1^4 + \frac{3}{4} (n_{12}v_1) (n_{12}v_2)
(v_1v_2) - 2 v_1^2 (v_1v_2) + \frac{1}{8} (v_1v_2)^2 + \frac{15}{16}
v_1^2 v_2^2\bigg) \nonumber \\ &+& G m_1 m_2 \bigg(-\frac{7}{4} (a_1
v_2) (n_{12}v_2) - \frac{1}{8} (n_{12} a_1) (n_{12}v_2)^2 +
\frac{7}{8} (n_{12} a_1) v_2^2\bigg)+1 \leftrightarrow 2\;.
\end{eqnarray}
Finally, the 3PN-accurate piece of the harmonic-coordinate Lagrangian
depends also on accelerations; it is notable that accelerations are
sufficient, there is no need to include, at the 3PN order, derivatives
of accelerations. We find \cite{ABF}

\begin{eqnarray}\label{14}
L_{\rm 3PN} &=& \frac{5}{128}m_1 v_1^8 + \frac{G^2 m_1^2
m_2}{r_{12}^2} \bigg(\frac{13}{18} (n_{12}v_1)^4 + \frac{83}{18}
(n_{12}v_1)^3 (n_{12}v_2) - \frac{35}{6} (n_{12}v_1)^2 (n_{12}v_2)^2
\nonumber\\ &-& \frac{245}{24} (n_{12}v_1)^2 v_1^2 + \frac{179}{12}
(n_{12}v_1) (n_{12}v_2) v_1^2 - \frac{235}{24} (n_{12}v_2)^2 v_1^2 +
\frac{373}{48} v_1^4 + \frac{529}{24} (n_{12}v_1)^2 (v_1v_2) \nonumber
\\ &-& \frac{97}{6} (n_{12}v_1) (n_{12}v_2) (v_1v_2) - \frac{719}{24}
v_1^2 (v_1v_2) + \frac{463}{24} (v_1v_2)^2 - \frac{7}{24}
(n_{12}v_1)^2 v_2^2 \nonumber \\ &-& \frac{1}{2} (n_{12}v_1)
(n_{12}v_2) v_2^2 + \frac{1}{4} (n_{12}v_2)^2 v_2^2 + \frac{463}{48}
v_1^2 v_2^2 - \frac{19}{2} (v_1v_2) v_2^2 + \frac{45}{16} v_2^4\bigg)
\nonumber \\ &+& G m_1 m_2 \bigg(\frac{3}{8} (a_1 v_2) (n_{12}v_1)
(n_{12}v_2)^2 + \frac{5}{12} (a_1 v_2) (n_{12}v_2)^3 + \frac{1}{8}
(n_{12} a_1) (n_{12}v_1) (n_{12}v_2)^3 \nonumber \\ &+& \frac{1}{16}
(n_{12} a_1) (n_{12}v_2)^4 + \frac{11}{4} (a_1 v_1) (n_{12}v_2) v_1^2
- (a_1 v_2) (n_{12}v_2) v_1^2 - 2 (a_1 v_1) (n_{12}v_2) (v_1v_2)
\nonumber \\ &+& \frac{1}{4} (a_1 v_2) (n_{12}v_2) (v_1v_2) +
\frac{3}{8} (n_{12} a_1) (n_{12}v_2)^2 (v_1v_2) - \frac{5}{8} (n_{12}
a_1) (n_{12}v_1)^2 v_2^2 \nonumber \\ &+& \frac{15}{8} (a_1 v_1)
(n_{12}v_2) v_2^2 - \frac{15}{8} (a_1 v_2) (n_{12}v_2) v_2^2 -
\frac{1}{2} (n_{12} a_1) (n_{12}v_1) (n_{12}v_2) v_2^2 \nonumber \\
&-& \frac{5}{16} (n_{12} a_1) (n_{12}v_2)^2 v_2^2\bigg) + \frac{G^2
m_1^2 m_2}{r_{12}} \bigg(-\frac{235}{24} (a_2 v_1) (n_{12}v_1) -
\frac{29}{24} (n_{12} a_2) (n_{12}v_1)^2 \nonumber \\ &-&
\frac{235}{24} (a_1 v_2) (n_{12}v_2) - \frac{17}{6} (n_{12} a_1)
(n_{12}v_2)^2 + \frac{185}{16} (n_{12} a_1) v_1^2 - \frac{235}{48}
(n_{12} a_2) v_1^2 \nonumber \\ &-& \frac{185}{8} (n_{12} a_1)
(v_1v_2) + \frac{20}{3} (n_{12} a_1) v_2^2\bigg) +
\frac{G m_1 m_2}{r_{12}} \bigg(-\frac{5}{32} (n_{12}v_1)^3
(n_{12}v_2)^3 \nonumber \\ &+& \frac{1}{8} (n_{12}v_1) (n_{12}v_2)^3
v_1^2 + \frac{5}{8} (n_{12}v_2)^4 v_1^2 - \frac{11}{16} (n_{12}v_1)
(n_{12}v_2) v_1^4 + \frac{1}{4} (n_{12}v_2)^2 v_1^4 + \frac{11}{16}
v_1^6 \nonumber\\ &-& \frac{15}{32} (n_{12}v_1)^2 (n_{12}v_2)^2
(v_1v_2) + (n_{12}v_1) (n_{12}v_2) v_1^2 (v_1v_2) + \frac{3}{8}
(n_{12}v_2)^2 v_1^2 (v_1v_2) \nonumber\\ &-& \frac{13}{16} v_1^4
(v_1v_2) + \frac{5}{16} (n_{12}v_1) (n_{12}v_2) (v_1v_2)^2 +
\frac{1}{16} (v_1v_2)^3 - \frac{5}{8} (n_{12}v_1)^2 v_1^2 v_2^2
\nonumber\\ &-& \frac{23}{32} (n_{12}v_1) (n_{12}v_2) v_1^2 v_2^2 +
\frac{1}{16} v_1^4 v_2^2 - \frac{1}{32} v_1^2 (v_1v_2) v_2^2\bigg)
\nonumber\\ &-&\frac{3}{8}\frac{G^4 m_1^4 m_2}{r_{12}^4 } + \frac{G^4
m_1^3 m_2^2}{r_{12}^4} \bigg(-\frac{5809}{280} + \frac{11}{3} \lambda
+ \frac{22}{3} \ln \left(\frac{r_{12}}{r'_1} \right)\bigg)
\nonumber\end{eqnarray}
\begin{eqnarray}
\qquad 
&+& \frac{G^3 m_1^2 m_2^2}{r_{12}^3} \bigg(\frac{383}{24}
(n_{12}v_1)^2 - \frac{889}{48} (n_{12}v_1) (n_{12}v_2) -
\frac{305}{72} v_1^2 +\frac{439}{144} (v_1v_2) \nonumber\\ &-&
\frac{123}{64} (n_{12}v_1)^2 \pi^2 + \frac{123}{64} (n_{12}v_1)
(n_{12}v_2) \pi^2 + \frac{41}{64} v_1^2\pi^2 - \frac{41}{64}
(v_1v_2)\pi^2\bigg) \nonumber\\ &+& \frac{G^3 m_1^3 m_2}{r_{12}^3}
\bigg(-\frac{8243}{210} (n_{12}v_1)^2 + \frac{15541}{420} (n_{12}v_1)
(n_{12}v_2) + \frac{3}{2} (n_{12}v_2)^2 \nonumber\\ &+&
\frac{15611}{1260} v_1^2 - \frac{17501}{1260} (v_1v_2) + \frac{5}{4}
v_2^2 + 22 (n_{12}v_1)(n_{12}v_{12}) \ln
\left(\frac{r_{12}}{r'_1} \right) \nonumber\\ &-& \frac{22}{3} (v_1v_{12}) \ln
\left(\frac{r_{12}}{r'_1} \right) \bigg)+1 \leftrightarrow 2\;.
\end{eqnarray}
See Ref. \cite{ABF} for the explicit expressions of the ten conserved
quantities, at the 3PN order, corresponding to the integrals of
energy, linear and angular momentum, and center-of-mass position.


%
\end{document}